\documentclass[showpacs,aps,prd]{revtex4}
\input epsf

\begin{document}
~~~~~~~~~~~~~~~~~~~~~~~~~~~~~~~~~~~~~~~~~~~~~~~~~~~~~~~~~
~~~~~~~~~~~~~~~~~~~~~~~~~~~~~~~~~~~~~~~~~~~~~~~~~~\texttt{\emph{talk
given at NSTAR 2009}}

\title{Heavy flavor baryon spectra via QCD sum rules}
\author{Jian-Rong Zhang and Ming-Qiu Huang}
\affiliation{Department of Physics, National University of Defense
Technology, Hunan 410073, China}

\begin{abstract}
In this talk, we give a short review of our recent works on studying
the singly heavy baryon, doubly heavy baryon, and triply heavy
baryon spectra from QCD sum rules.
\end{abstract}
\maketitle

\section{Introduction}\label{sec1}
The heavy baryon is a exciting and remarkable topic nowadays.
Experimentally, the field of heavy hadron spectroscopy is
experiencing a rapid advancement and plenty of heavy baryons have
already been observed up to now \cite{experiments,PDG}. The
feasibility of doubly and triply heavy baryons investigated at the
Large Hadron Collider (with the design luminosity values of
$\mathcal{L}=10^{34}~\mbox{cm}^{-2}\mbox{s}^{-1}$ and
$\sqrt{s}=14~\mbox{TeV}$) was presented in some works, for instance,
Refs. \cite{production,production1}. Theoretically, various models
have been utilized to compute heavy baryon masses, such as quark
models \cite{quark model,quark model 1,quark model
2,Martynenko,Hasenfratz,Bjorken}, mass formulas \cite{mass
formular,mass formulas}, lattice QCD stimulations
\cite{lattice,lat}, and other approaches \cite{MIT bag
model,YuJia,renewed}. One can also resort to a vigorous and reliable
working tool in hadron physics, the QCD sum rules, which are still
being actively used judging by the near $3500$ and growing citations
of the seminal papers \cite{svzsum} of M.~A.~Shifman,
A.~I.~Vainshtein, and V.~I.~Zakharov. The method is a
nonperturbative analytic formalism firmly entrenched in QCD (for
reviews see \cite{svz,svz1} and references therein). QCD sum rules
for baryons \cite{Ioffe} suggested by B.~L.~Ioffe generalize the
method from the mesonic states to the baryonic cases. With QCD sum
rules, heavy baryon masses were primarily calculated by
E.~V.~Shuryak in heavy quark limit \cite{evs}, and subsequently in
the Heavy Quark Effective Theory by some theorists, for example,
A.~G.~Grozin, Y.~B.~Dai, S.~Groote etc.
\cite{alfa,radiative,HQET,zhu,mqhuang}. There also have many works
been done basing on the full theory by E.~Bagan, V.~V.~Kiselev,
T.~M.~Aliev, M.~Nielsen etc. \cite{NRQCDSR,Bagan,full,Azizi}, as
well as our studies on singly \cite{our,our1}, doubly \cite{our2},
and triply heavy baryon spectra \cite{our3} from QCD sum rules.
Presently, we would like to briefly review our those works and make
some discussions.

The content of the review is as follows. In Sec. \ref{sec2}, the
main results are collected in comparison with experimental data and
other approaches, followed by some discussions. Section \ref{sec3}
contains a concise summary and outlook.
\section{Heavy baryons in QCD sum rules}\label{sec2}
The QCD sum rule approach devotes to bridge the gap between the
perturbative and nonperturbative sectors by employing the language
of dispersion relations and represents an attempt to link the hadron
phenomenology with the interactions of quarks and gluons. There are
three leading ingredients for this method: a phenomenological
description of the correlator, a theoretical description of the same
correlator via an operator product expansion (OPE), and a procedure
for matching these two descriptions and extracting the parameters
that characterize the hadronic state of interest. Meanwhile, the QCD
sum rule accuracy is limited by a very complicated hadronic
dispersion integrals and by the approximations in the OPE of the
correlator. The basic point of this method is the choice of
appropriate interpolating current. In a tentative diquark-quark
picture for the singly heavy baryon $qqQ$ system, the $Q$ orbits the
$qq$ pair. For the ground states, the currents are correlated with
the spin-parity quantum numbers $0^{+}$ and $1^{+}$ for the $qq$
diquark system, along with the heavy quark $Q$ forming the state
with $J^{P}=\frac{1}{2}^{+}$ and the pair of degenerate states. For
the latter case, the $qq$ diquark has spin $1$, and the spin of the
third quark is either parallel, $J^{P}=\frac{3}{2}^{+}$, or
antiparallel, $J^{P}=\frac{1}{2}^{+}$, to the diquark. Similarly,
one could assume the $(QQ)-q$ configuration for doubly heavy baryon
$QQq$ and $(QQ)-Q'$ for triply heavy baryon $QQQ'$, respectively.
Thereby, we principally adopt the similar forms of Ioffe currents
discussed minutely in Refs. \cite{evs,Ioffe}, with
\begin{eqnarray}
j_{\Lambda_{Q}}&=&\varepsilon_{abc}(q_{1a}^{T}C\Gamma_{k}q_{2b})\Gamma_{k}^{'}Q_{c},\nonumber\\
j_{\Lambda_{1Q}}&=&\varepsilon_{abc}(q_{1a}^{T}C\Gamma_{k}q_{2b})\Gamma_{k}^{'}Q_{c},\nonumber\\
j_{\Lambda_{1Q}^{*}}&=&\varepsilon_{abc}\frac{1}{\sqrt{3}}[2(q_{1a}^{T}C\Gamma_{k}Q_{b})\Gamma_{k}^{'}q_{2c}
+(q_{1a}^{T}C\Gamma_{k}q_{2b})\Gamma_{k}^{'}Q_{c}],\nonumber\\
j_{\Sigma_{Q}}&=&\varepsilon_{abc}(q_{1a}^{T}C\Gamma_{k}q_{2b})\Gamma_{k}^{'}Q_{c},\nonumber\\
j_{\Sigma_{Q}^{*}}&=&\varepsilon_{abc}\frac{1}{\sqrt{3}}[2(q_{1a}^{T}C\Gamma_{k}Q_{b})\Gamma_{k}^{'}q_{2c}
+(q_{1a}^{T}C\Gamma_{k}q_{2b})\Gamma_{k}^{'}Q_{c}],\nonumber\\
j_{\Xi_{Q}}&=&\varepsilon_{abc}(q_{1a}^{T}C\Gamma_{k}s_{b})\Gamma_{k}^{'}Q_{c},\nonumber\\
j_{\Xi_{1Q}}&=&\varepsilon_{abc}(q_{1a}^{T}C\Gamma_{k}s_{b})\Gamma_{k}^{'}Q_{c},\nonumber\\
j_{\Xi_{1Q}^{*}}&=&\varepsilon_{abc}\frac{1}{\sqrt{3}}[(q_{1a}^{T}C\Gamma_{k}Q_{b})\Gamma_{k}^{'}s_{c}
+(q_{1a}^{T}C\Gamma_{k}s_{b})\Gamma_{k}^{'}Q_{c}],\nonumber\\
j_{\Xi_{Q}^{'}}&=&\varepsilon_{abc}(q_{1a}^{T}C\Gamma_{k}s_{b})\Gamma_{k}^{'}Q_{c},\nonumber\\
j_{\Xi_{Q}^{'*}}&=&\varepsilon_{abc}\frac{1}{\sqrt{3}}[2(q_{1a}^{T}C\Gamma_{k}Q_{b})\Gamma_{k}^{'}s_{c}
+(q_{1a}^{T}C\Gamma_{k}s_{b})\Gamma_{k}^{'}Q_{c}],\nonumber\\
j_{\Omega_{Q}}&=&\varepsilon_{abc}(s^{T}_{a}C\Gamma_{k}s_{b})\Gamma_{k}^{'}Q_{c},\nonumber\\
j_{\Omega_{Q}^{*}}&=&\varepsilon_{abc}\frac{1}{\sqrt{3}}[2(s^{T}_{a}C\Gamma_{k}Q_{b})\Gamma_{k}^{'}s_{c}
+(s^{T}_{a}C\Gamma_{k}s_{b})\Gamma_{k}^{'}Q_{c}],\nonumber
\end{eqnarray}
for singly heavy baryons,
\begin{eqnarray}
j_{\Xi_{QQ}}&=&\varepsilon_{abc}(Q_{a}^{T}C\Gamma_{k}Q_{b})\Gamma_{k}^{'}q_{c},\nonumber\\
j_{\Xi_{QQ}^{*}}&=&\varepsilon_{abc}\frac{1}{\sqrt{3}}[2(q_{a}^{T}C\Gamma_{k}Q_{b})\Gamma_{k}^{'}Q_{c}+(Q_{a}^{T}C\Gamma_{k}Q_{b})\Gamma_{k}^{'}q_{c}],\nonumber\\
j_{\Omega_{QQ}}&=&\varepsilon_{abc}(Q_{a}^{T}C\Gamma_{k}Q_{b})\Gamma_{k}^{'}s_{c},\nonumber\\
j_{\Omega_{QQ}^{*}}&=&\varepsilon_{abc}\frac{1}{\sqrt{3}}[2(s_{a}^{T}C\Gamma_{k}Q_{b})\Gamma_{k}^{'}Q_{c}+(Q_{a}^{T}C\Gamma_{k}Q_{b})\Gamma_{k}^{'}s_{c}],\nonumber\\
j_{\Xi_{QQ'}}&=&\varepsilon_{abc}(Q_{a}^{T}C\Gamma_{k}Q'_{b})\Gamma_{k}^{'}q_{c},\nonumber\\
j_{\Xi_{QQ'}^{*}}&=&\varepsilon_{abc}\frac{1}{\sqrt{3}}[(q_{a}^{T}C\Gamma_{k}Q_{b})\Gamma_{k}^{'}Q'_{c}+(q_{a}^{T}C\Gamma_{k}Q'_{b})\Gamma_{k}^{'}Q_{c}+(Q_{a}^{T}C\Gamma_{k}Q_{b})\Gamma_{k}^{'}q_{c}],\nonumber\\
j_{\Omega_{QQ'}}&=&\varepsilon_{abc}(Q_{a}^{T}C\Gamma_{k}Q'_{b})\Gamma_{k}^{'}s_{c},\nonumber\\
j_{\Omega_{QQ'}^{*}}&=&\varepsilon_{abc}\frac{1}{\sqrt{3}}[(s_{a}^{T}C\Gamma_{k}Q_{b})\Gamma_{k}^{'}Q'_{c}+(s_{a}^{T}C\Gamma_{k}Q'_{b})\Gamma_{k}^{'}Q_{c}+(Q_{a}^{T}C\Gamma_{k}Q_{b})\Gamma_{k}^{'}s_{c}],\nonumber\\
j_{\Xi'_{QQ'}}&=&\varepsilon_{abc}(Q_{a}^{T}C\Gamma_{k}Q'_{b})\Gamma_{k}^{'}q_{c},\nonumber\\
j_{\Omega'_{QQ'}}&=&\varepsilon_{abc}(Q_{a}^{T}C\Gamma_{k}Q'_{b})\Gamma_{k}^{'}s_{c},\nonumber
\end{eqnarray}
for doubly heavy baryons, and
\begin{eqnarray}
j_{\Omega_{QQQ}}&=&\varepsilon_{abc}(Q_{a}^{T}C\Gamma_{k}Q_{b})\Gamma_{k}^{'}Q_{c},\nonumber\\
j_{\Omega_{QQQ'}}&=&\varepsilon_{abc}(Q_{a}^{T}C\Gamma_{k}Q_{b})\Gamma_{k}^{'}Q'_{c},\nonumber\\
j_{\Omega_{QQQ'}^{*}}&=&\varepsilon_{abc}\frac{1}{\sqrt{3}}[2(Q_{a}^{T}C\Gamma_{k}Q'_{b})\Gamma_{k}^{'}Q_{c}+(Q_{a}^{T}C\Gamma_{k}Q_{b})\Gamma_{k}^{'}Q'_{c}],\nonumber\\
j_{\Omega_{QQQ'}'}&=&\varepsilon_{abc}(Q_{a}^{T}C\Gamma_{k}Q_{b})\Gamma_{k}^{'}Q'_{c},\nonumber
\end{eqnarray}
for triply heavy baryons. Here the index $T$ means matrix
transposition, $C$ is the charge conjugation matrix, $a$, $b$, and
$c$ are color indices, $Q$ and $Q'$ denote heavy quarks, and $q$ is
$u$ or $d$. The choice of $\Gamma_{k}$ and $\Gamma_{k}^{'}$ matrices
are listed in TABLE \ref{table:1}.
\begin{table}[htb!]\caption{The choice of $\Gamma_{k}$ and $\Gamma_{k}^{'}$ matrices in baryonic currents. The index $d$ in $S_{d}$, $L_{d}$, and $J_{d}^{P_{d}}$ means diquark. $\{..\}$ denotes the diquark in the axial
vector state, and $[..]$ denotes the diquark in the scalar state.}
 \centerline{\begin{tabular}{ p{1.5cm} p{2.5cm} p{2.0cm} p{1cm} p{1cm} p{2.0cm} p{2.0cm} p{2.0cm}} \hline\hline
Baryon              & quark content        &$J^{P}$               &  $S_{d}$     &  $L_{d}$     &  $J_{d}^{P_{d}}$         &   $\Gamma_{k}$     &     $\Gamma_{k}^{'}$           \\
\hline
$\Lambda_{Q}$       &$[qq]Q$               &$\frac{1}{2}^{+} $    &      0       &      0       &        $0^{+}$           &   $\gamma_{5}$     &     $1$                        \\
$\Lambda_{1Q}$      &$[qq]Q$               &$\frac{1}{2}^{-}$     &      0       &      1       &        $1^{-}$           &   $\gamma_{5}$     &     $\gamma_{\mu}$             \\
$\Lambda_{1Q}^{*}$  &$[qq]Q$               &$\frac{3}{2}^{-}$     &      0       &      1       &        $1^{-}$           &   $\gamma_{5}$     &     $\gamma_{\mu}$             \\
$\Sigma_{Q}$        &$\{qq\}Q$             &$\frac{1}{2}^{+} $    &      1       &      0       &        $1^{+}$           &   $\gamma_{\mu}$   &     $\gamma_{\mu}\gamma_{5}$   \\
$\Sigma_{Q}^{*}$    &$\{qq\}Q$             &$\frac{3}{2}^{+} $    &      1       &      0       &        $1^{+}$           &   $\gamma_{\mu}$   &     $\gamma_{\mu}\gamma_{5}$   \\
$\Xi_{Q}$           &$[qs]Q$               &$\frac{1}{2}^{+} $    &      0       &      0       &        $0^{+}$           &   $\gamma_{5}$     &     $1$                        \\
$\Xi_{1Q}$          &$[qs]Q$               &$\frac{1}{2}^{-}$     &      0       &      1       &        $1^{-}$           &   $\gamma_{5}$     &     $\gamma_{\mu}$             \\
$\Xi_{1Q}^{*}$      &$[qs]Q$               &$\frac{3}{2}^{-}$     &      0       &      1       &        $1^{-}$           &   $\gamma_{5}$     &     $\gamma_{\mu}$             \\
$\Xi_{Q}^{'}$       &$\{qs\}Q$             &$\frac{1}{2}^{+} $    &      1       &      0       &        $1^{+}$           &   $\gamma_{\mu}$   &     $\gamma_{\mu}\gamma_{5}$   \\
$\Xi_{Q}^{'*}$      &$\{qs\}Q$             &$\frac{3}{2}^{+} $    &      1       &      0       &        $1^{+}$           &   $\gamma_{\mu}$   &     $\gamma_{\mu}\gamma_{5}$   \\
$\Omega_{Q}$        &$\{ss\}Q$             &$\frac{1}{2}^{+} $    &      1       &      0       &        $1^{+}$           &   $\gamma_{\mu}$   &     $\gamma_{\mu}\gamma_{5}$   \\
$\Omega_{Q}^{*}$    &$\{ss\}Q$             &$\frac{3}{2}^{+} $    &      1       &      0       &        $1^{+}$           &   $\gamma_{\mu}$   &     $\gamma_{\mu}\gamma_{5}$   \\
\hline
$\Xi_{QQ}$          &$\{QQ\}q$             &$\frac{1}{2}^{+} $    &      1       &      0       &        $1^{+}$           &   $\gamma_{\mu}$   &     $\gamma_{\mu}\gamma_{5}$   \\
$\Xi_{QQ}^{*}$      &$\{QQ\}q$             &$\frac{3}{2}^{+} $    &      1       &      0       &        $1^{+}$           &   $\gamma_{\mu}$   &     $1$                        \\
$\Omega_{QQ}$       &$\{QQ\}s$             &$\frac{1}{2}^{+} $    &      1       &      0       &        $1^{+}$           &   $\gamma_{\mu}$   &     $\gamma_{\mu}\gamma_{5}$   \\
$\Omega_{QQ}^{*}$   &$\{QQ\}s$             &$\frac{3}{2}^{+} $    &      1       &      0       &        $1^{+}$           &   $\gamma_{\mu}$   &     $1$                        \\
$\Xi_{QQ'}$         &$\{QQ'\}q$            &$\frac{1}{2}^{+} $    &      1       &      0       &        $1^{+}$           &   $\gamma_{\mu}$   &     $\gamma_{\mu}\gamma_{5}$   \\
$\Xi_{QQ'}^{*}$     &$\{QQ'\}q$            &$\frac{3}{2}^{+} $    &      1       &      0       &        $1^{+}$           &   $\gamma_{\mu}$   &     $1$                        \\
$\Omega_{QQ'}$      &$\{QQ'\}s$            &$\frac{1}{2}^{+} $    &      1       &      0       &        $1^{+}$           &   $\gamma_{\mu}$   &     $\gamma_{\mu}\gamma_{5}$   \\
$\Omega_{QQ'}^{*}$  &$\{QQ'\}s$            &$\frac{3}{2}^{+} $    &      1       &      0       &        $1^{+}$           &   $\gamma_{\mu}$   &     $1$                        \\
$\Xi_{QQ'}^{'}$     &$[QQ']q$              &$\frac{1}{2}^{+} $    &      0       &      0       &        $0^{+}$           &   $\gamma_{5}$     &     $1$                        \\
$\Omega_{QQ'}^{'}$  &$[QQ']s$              &$\frac{1}{2}^{+} $    &      0       &      0       &        $0^{+}$           &   $\gamma_{5}$     &     $1$                        \\
\hline
$\Omega_{QQQ}$      &$\{QQ\}Q$             &$\frac{3}{2}^{+} $    &      1       &      0       &        $1^{+}$           &   $\gamma_{\mu}$   &     $1$                        \\
$\Omega_{QQQ'}$     &$\{QQ\}Q'$            &$\frac{1}{2}^{+} $    &      1       &      0       &        $1^{+}$           &   $\gamma_{\mu}$   &     $\gamma_{\mu}\gamma_{5}$   \\
$\Omega_{QQQ'}^{*}$ &$\{QQ\}Q'$            &$\frac{3}{2}^{+} $    &      1       &      0       &        $1^{+}$           &   $\gamma_{\mu}$   &     $1$                        \\
$\Omega_{QQQ'}'$    &$[QQ]Q'$              &$\frac{1}{2}^{+} $    &      0       &      0       &        $0^{+}$           &   $\gamma_{5}$     &     $1$                        \\
\hline\hline
\end{tabular}}
\label{table:1}
\end{table}

Concretely, coming down to the mass sum rules for the singly heavy
baryon as an example, the starting point is the two-point correlator
\begin{eqnarray}\label{correlator}
\Pi(q^{2})=i\int
d^{4}x\mbox{e}^{iq.x}\langle0|T[j(x)\overline{j}(0)]|0\rangle.
\end{eqnarray}
Lorentz covariance implies that the correlator (\ref{correlator})
has the form
\begin{eqnarray}
\Pi(q^{2})=\rlap/q\Pi_{1}(q^{2})+\Pi_{2}(q^{2}).
\end{eqnarray}
For each invariant function $\Pi_{1}$ and $\Pi_{2}$, a sum rule can
be obtained.

In the phenomenology side, the correlator can be expressed as a
dispersion integral over a physical spectral function
\begin{eqnarray}
\Pi(q^{2})=\lambda^{2}_H\frac{\rlap/q+M_{H}}{M_{H}^{2}-q^{2}}+\frac{1}{\pi}\int_{s_{0}}
^{\infty}ds\frac{\mbox{Im}\Pi^{\mbox{phen}}(s)}{s-q^{2}}+\mbox{subtractions},
\end{eqnarray}
where $M_{H}$ denotes the heavy baryon mass.

In the OPE side, the correlator can be written in terms of a
dispersion relation as
\begin{eqnarray}
\Pi_{i}(q^{2})=\int_{m_{Q}^{2}}^{\infty}ds\frac{\rho_{i}(s)}{s-q^{2}},~~i=1,2.
\end{eqnarray}

After equating the two sides, assuming quark-hadron duality, making
a Borel transform, and eliminating the baryon coupling constant
$\lambda_H$, the sum rules can be written as,
\begin{eqnarray}
M_{H}^{2}&=&\int_{m_{Q}^{2}}^{s_{0}}ds\rho_{i}(s)s e^{-s/M^{2}}/
\int_{m_{Q}^{2}}^{s_{0}}ds\rho_{i}(s)e^{-s/M^{2}},~~i=1,2.
\end{eqnarray}

For brevity, more detailed descriptions of the calculation
procedures will not be iterated here, which can be found in Refs.
\cite{our,our1,our2,our3}. The final results are collected together
with the available experimental data and other theoretical
predictions in Tables II-IV. It is worth noting that uncertainty in
our results are merely due to the sum rule windows, not involving
the ones rooting in the variation of the quark masses and QCD
parameters. Note that the QCD $O(\alpha_s)$ corrections are not
covered in these works. However, it is expected that the QCD
$O(\alpha_s)$ corrections might be under control since a partial
cancelation occurs in the ratio obtaining the mass sum rules. This
has been proved to be true in the analysis for the singly heavy
baryons (the radiative corrections to the perturbative terms
increase the calculated baryon masses by about $10\%$) in Ref.
\cite{radiative} and for the heavy mesons (the value of $f_{D}$
increases by $12\%$ after the inclusion of the $O(\alpha_s)$
correction) in Ref. \cite{Narison}. Although the mass values for
doubly heavy baryons are consistent with other theoretical
predictions, some of the absolute differences from them are not
small, for instance, the masses of $\Xi_{cc}$, $\Omega_{cc}$, and
$\Xi_{cb}^{*}$, whereas, the relative discrepancies are in the
tolerable ranges of the sum rule accuracy. Visually, the Borel
curves for $\Xi_{cc}$, $\Omega_{cc}$, and $\Xi_{cb}^{*}$ are not
very flat, but it is difficult to find much better sum rule windows.
That's probably because the condensate contributions for them, which
may play an important role in stabilizing the Borel curves, nearly
vanished or are small. The stability of those three curves might be
improved by including some higher dimension condensate
contributions. For triply heavy baryons, one can find that our
central values are lower than potential model predictions, in
particular, for $\Omega_{bbb}$, slightly more than $\mbox{1~GeV}$,
whereas the relative discrepancy approximates to $10\%$, which is
still acceptable. In addition, our result for $\Omega_{ccc}$ agrees
well with the lattice QCD value in Ref. \cite{lat}, but the other
comparisons for triply heavy baryons cannot be made for the absence
of relevant lattice results by this time.

\begin{table}\caption{ The mass spectra of charmed and bottom baryons (mass in
unit of$~\mbox{MeV}$ except for ``Our works").}
 \centerline{\begin{tabular}{ c  c  c  c  c  c  c  c c c c}  \hline\hline
Baryon                          &     $J^{P}$            &  $S_{\ell}$  &  $L_{\ell}$  &   $J_{\ell}^{P_{\ell}}$  &  Experiments  \cite{experiments,PDG}                                         & Our works~(\mbox{GeV}) \cite{our,our1}               & Refs. \cite{quark model}                        & Ref. \cite{mass formular}               & Ref. \cite{lattice}     & Ref. \cite{zhu}          \\
\hline
$\Lambda_{c}^{+}$               &  $\frac{1}{2}^{+} $    &     0        &      0       &        $0^{+}$           & $2286.46\pm0.14$                                                             & $2.31\pm0.19$                                        & 2297                                            &   2285                                  &    2290                 & $2271_{-49}^{+67}$       \\
\hline
$\Lambda_{c}(2593)^{+}$         &  $\frac{1}{2}^{-} $    &     0        &      1       &        $1^{-}$           & $2595.4\pm0.6$                                                               & $2.53\pm0.22$                                        & 2598                                            &                                         &                         &                          \\
\hline
$\Lambda_{c}(2625)^{+}$         &  $\frac{3}{2}^{-} $    &     0        &      1       &        $1^{-}$           & $2628.1\pm0.6$                                                               & $2.58\pm0.24$                                        & 2628                                            &                                         &                         &                          \\
\hline
$\Sigma_{c}(2455)^{0}$          &  $\frac{1}{2}^{+} $    &     1        &      0       &        $1^{+}$           & $2453.76\pm0.18$                                                             & $2.40\pm0.31$                                        & 2439                                            &   2453                                  &    2452                 & $2411_{-81}^{+93}$       \\
\hline
$\Sigma_{c}(2520)^{0}$          &  $\frac{3}{2}^{+} $    &     1        &      0       &        $1^{+}$           & $2518.0\pm0.5$                                                               & $2.56\pm0.24$                                        & 2518                                            &   2520                                  &    2538                 & $2534_{-81}^{+96}$       \\
\hline
$\Xi_{c}^{0}$                   &  $\frac{1}{2}^{+} $    &     0        &      0       &        $0^{+}$           & $2471.0\pm0.4$                                                               & $2.48\pm0.21$                                        & 2481                                            &   2468                                  &    2473                 & $2432_{-68}^{+79}$       \\
\hline
$\Xi_{c}(2790)^{0}$             &  $\frac{1}{2}^{-} $    &     0        &      1       &        $1^{-}$           & $2791.9\pm3.3$                                                               & $2.65\pm0.27$                                        & 2801                                            &                                         &                         &                          \\
\hline
$\Xi_{c}(2815)^{0}$             &  $\frac{3}{2}^{-}$     &     0        &      1       &        $1^{-}$           & $2818.2\pm2.1$                                                               & $2.69\pm0.29$                                        & 2820                                            &                                         &                         &                          \\
\hline
$\Xi_{c}'^{0}$                  &  $\frac{1}{2}^{+} $    &     1        &      0       &        $1^{+}$           & $2578.0\pm2.9$                                                               & $2.50\pm0.29$                                        & 2578                                            &   2580                                  &    2599                 & $2508_{-91}^{+97}$       \\
\hline
$\Xi_{c}(2645)^{0}$             &  $\frac{3}{2}^{+} $    &     1        &      0       &        $1^{+}$           & $2646.1\pm1.2$                                                               & $2.64\pm0.22$                                        & 2654                                            &   2650                                  &    2680                 & $2634_{-94}^{+102}$      \\
\hline
$\Omega_{c}^{0}$                &   $\frac{1}{2}^{+}$    &     1        &      0       &        $1^{+}$           & $2697.5\pm2.6$                                                               & $2.62\pm0.29$                                        & 2698                                            &   2710                                  &    2678                 & $2657_{-99}^{+102}$      \\
\hline
$\Omega_{c}(2768)^{0}$          &  $\frac{3}{2}^{+} $    &     1        &      0       &        $1^{+}$           & $2768.3\pm3.0$                                                               & $2.74\pm0.23$                                        & 2768                                            &   2770                                  &    2752                 & $2790_{-105}^{+109}$     \\
\hline
$\Lambda_{b}$                   &  $\frac{1}{2}^{+} $    &     0        &      0       &        $0^{+}$           & $5619.7\pm1.2$                                                               & $5.69\pm0.13$                                        & 5622                                            &   5620                                  &    5672                 & $5637_{-56}^{+68}$       \\
                                &                        &              &              &                          & $5624\pm9$                                                                   &                                                      &                                                 &                                         &                         &                          \\
\hline
$\Lambda_{1b}$                  &  $\frac{1}{2}^{-} $    &     0        &      1       &        $1^{-}$           &                                                                              & $5.85\pm0.15$                                        & 5930                                            &                                         &                         &                          \\
\hline
$\Lambda_{1b}^{*}$              &  $\frac{3}{2}^{-} $    &     0        &      1       &        $1^{-}$           &                                                                              & $5.90\pm0.16$                                        & 5947                                            &                                         &                         &                          \\
\hline
$\Sigma_{b}^{+}$                &  $\frac{1}{2}^{+} $    &     1        &      0       &        $1^{+}$           & $5807.8_{-2.2}^{+2.0}\pm1.7$                                                 & $5.73\pm0.21$                                        & 5805                                            &   5820                                  &    5847                 & $5809_{-76}^{+82}$       \\
$\Sigma_{b}^{-}$                &                        &              &              &                          & $5815.2\pm1.0\pm1.7$                                                         &                                                      &                                                 &                                         &                         &                          \\
\hline
$\Sigma_{b}^{*+}$               &  $\frac{3}{2}^{+} $    &     1        &      0       &        $1^{+}$           & $5829.0_{-1.8}^{+1.6}$$_{-1.8}^{+1.7}$                                       & $5.81\pm0.19$                                        & 5834                                            &   5850                                  &    5871                 & $5835_{-77}^{+82}$       \\
$\Sigma_{b}^{*-}$               &                        &              &              &                          & $5836.4\pm2.0_{-1.7}^{+1.8}$                                                 &                                                      &                                                 &                                         &                         &                          \\
\hline
$\Xi_{b}^{0}$                   &  $\frac{1}{2}^{+} $    &     0        &      0       &        $0^{+}$           &  $5792.9\pm2.5\pm1.7$                                                        & $5.75\pm0.13$                                        & 5812                                            &  5810                                   &    5788                 & $5780_{-68}^{+73}$       \\
\hline
$\Xi_{1b}$                      &  $\frac{1}{2}^{-} $    &     0        &      1       &        $1^{-}$           &                                                                              & $5.95\pm0.16$                                        & 6119                                            &                                         &                         &                          \\
\hline
$\Xi_{1b}^{*}$                  &  $\frac{3}{2}^{-}$     &     0        &      1       &        $1^{-}$           &                                                                              & $5.99\pm0.17$                                        & 6130                                            &                                         &                         &                          \\
\hline
$\Xi_{b}^{'}$                   &  $\frac{1}{2}^{+} $    &     1        &      0       &        $1^{+}$           &                                                                              & $5.87\pm0.20$                                        & 5937                                            &  5950                                   &    5936                 & $5903_{-79}^{+81}$       \\
\hline
$\Xi_{b}^{'*}$                  &  $\frac{3}{2}^{+} $    &     1        &      0       &        $1^{+}$           &                                                                              & $5.94\pm0.17$                                        & 5963                                            &  5980                                   &    5959                 & $5929_{-79}^{+83}$       \\
\hline
$\Omega_{b}$                    &  $\frac{1}{2}^{+}$     &     1        &      0       &        $1^{+}$           &  $6165\pm10\pm13$                                                            & $5.89\pm0.18$                                        & 6065                                            &  6060                                   &    6040                 & $6036\pm81$              \\
\hline
$\Omega_{b}^{*}$                &  $\frac{3}{2}^{+} $    &     1        &      0       &        $1^{+}$           &                                                                              & $6.00\pm0.16$                                        & 6090                                            &  6090                                   &    6060                 & $6063_{-82}^{+83}$       \\
\hline\hline
\end{tabular}} \label{table1}
\end{table}

\begin{table}\caption{ The mass spectra of doubly heavy baryons (mass in
unit of$~\mbox{GeV}$).}
 \centerline{\begin{tabular}{ c  c  c  c  c  c  c  c  c c c c }  \hline\hline
Baryon                          & content                &     $J^{P}$            &  $S_{d}$     &  $L_{d}$     &   $J_{d}^{P_{d}}$         & Our work \cite{our2}  &Ref. \cite{quark model 1}&Ref. \cite{mass formulas}& Ref. \cite{MIT bag model} & Refs. \cite{NRQCDSR} &  Refs. \cite{Bagan}\\
\hline
 $\Xi_{cc}$                     &$\{cc\}q$               &  $\frac{1}{2}^{+}$     &     1        &      0       &        $1^{+}$            & $4.26\pm0.19$         &    $3.620$              &     $3.676$             &  $3.520$                  &  $3.55\pm0.08$       &   $3.48\pm0.06$    \\
\hline
 $\Xi_{cc}^{*}$                 &$\{cc\}q$               &  $\frac{3}{2}^{+}$     &     1        &      0       &        $1^{+}$            & $3.90\pm0.10$         &    $3.727$              &     $3.746$             &  $3.63$                   &                      &   $3.58\pm0.05$    \\
\hline
 $\Omega_{cc}$                  &$\{cc\}s$               &  $\frac{1}{2}^{+}$     &     1        &      0       &        $1^{+}$            & $4.25\pm0.20$         &    $3.778$              &     $3.787$             &  $3.619$                  &  $3.65\pm0.07$       &                    \\
\hline
 $\Omega_{cc}^{*}$              &$\{cc\}s$               &  $\frac{3}{2}^{+}$     &     1        &      0       &        $1^{+}$            & $3.81\pm0.06$         &    $3.872$              &     $3.851$             &  $3.721$                  &                      &                    \\
\hline
 $\Xi_{bb}$                     &$\{bb\}q$               &  $\frac{1}{2}^{+}$     &     1        &      0       &        $1^{+}$            & $9.78\pm0.07$         &    $10.202$             &                         &  $10.272$                 &  $10.00\pm0.08$      &   $9.94\pm0.91$    \\
\hline
 $\Xi_{bb}^{*}$                 &$\{bb\}q$               &  $\frac{3}{2}^{+}$     &     1        &      0       &        $1^{+}$            & $10.35\pm0.08$        &    $10.237$             &     $10.398$            &  $10.337$                 &                      &   $10.33\pm1.09$   \\
\hline
 $\Omega_{bb}$                  &$\{bb\}s$               &  $\frac{1}{2}^{+}$     &     1        &      0       &        $1^{+}$            & $9.85\pm0.07$         &    $10.359$             &                         &  $10.369$                 & $10.09\pm0.07$       &                    \\
\hline
 $\Omega_{bb}^{*}$              &$\{bb\}s$               &  $\frac{3}{2}^{+}$     &     1        &      0       &        $1^{+}$            & $10.28\pm0.05$        &    $10.389$             &     $10.483$            &  $10.429$                 &                      &                    \\
 \hline
 $\Xi_{cb}$                     &$\{cb\}q$               &  $\frac{1}{2}^{+}$     &     1        &      0       &        $1^{+}$            & $6.75\pm0.05$         &    $6.933$              &     $7.053$             &  $6.838$                  & $6.79\pm0.08$        &                    \\
\hline
 $\Xi_{cb}^{*}$                 &$\{cb\}q$               &  $\frac{3}{2}^{+}$     &     1        &      0       &        $1^{+}$            & $8.00\pm0.26$         &    $6.980$              &     $7.083$             &  $6.986$                  &                      &                    \\
\hline
 $\Omega_{cb}$                  &$\{cb\}s$               &  $\frac{1}{2}^{+}$     &     1        &      0       &        $1^{+}$            & $7.02\pm0.08$         &    $7.088$              &     $7.148$             &  $6.941$                  & $6.89\pm0.07$        &                    \\
\hline
 $\Omega_{cb}^{*}$              &$\{cb\}s$               &  $\frac{3}{2}^{+}$     &     1        &      0       &        $1^{+}$            & $7.54\pm0.08$         &    $7.130$              &     $7.165$             &  $7.077$                  &                      &                    \\
\hline
 $\Xi_{cb}^{'}$                 &$[cb]q$                 &  $\frac{1}{2}^{+}$     &     0        &      0       &        $0^{+}$            & $6.95\pm0.08$         &    $6.963$              &     $7.062$             &  $7.028$                  &                      &   $6.44\pm0.19$    \\
\hline
 $\Omega_{cb}^{'}$              &$[cb]s$                 &  $\frac{1}{2}^{+}$     &     0        &      0       &        $0^{+}$            & $7.02\pm0.08$         &    $7.116$              &     $7.151$             &  $7.116$                  &                      &                    \\
\hline\hline
\end{tabular}} \label{table2}
\end{table}

\begin{table}\caption{ The mass spectra of triply heavy baryons (mass in
unit of$~\mbox{GeV}$).}
 \centerline{\begin{tabular}{ c c c c c c c c c c c c c}  \hline\hline
Baryon                          & quark content          &     $J^{P}$            &  $S_{d}$     &  $L_{d}$     &   $J_{d}^{P_{d}}$         & Our work \cite{our3} &Ref. \cite{Martynenko} &Ref. \cite{Hasenfratz} &Ref. \cite{Bjorken} &Ref. \cite{lat}      &Ref. \cite{YuJia}   \\
\hline
 $\Omega_{ccc}$                 &$\{cc\}c$               &  $\frac{3}{2}^{+}$     &     1        &      0       &        $1^{+}$            &   $4.67\pm0.15$      &   $4.803$             &    $4.79$             &    $4.925$         &     $4.681$         &   $4.76$           \\
\hline
 $\Omega_{bbb}$                 &$\{bb\}b$               &  $\frac{3}{2}^{+}$     &     1        &      0       &        $1^{+}$            &   $13.28\pm0.10$     &   $14.569$            &    $14.30$            &    $14.760$        &                     &   $14.37$          \\
 \hline
 $\Omega_{ccb}$                 &$\{cc\}b$               &  $\frac{1}{2}^{+}$     &     1        &      0       &        $1^{+}$            &   $7.41\pm0.13$      &   $8.018$             &                       &                    &                     &                    \\
\hline
 $\Omega_{ccb}^{*}$             &$\{cc\}b$               &  $\frac{3}{2}^{+}$     &     1        &      0       &        $1^{+}$            &   $7.45\pm0.16$      &   $8.025$             &    $8.03$             &    $8.200$         &                     &   $7.98$           \\
\hline
 $\Omega_{bbc}$                 &$\{bb\}c$               &  $\frac{1}{2}^{+}$     &     1        &      0       &        $1^{+}$            &   $10.30\pm0.10$     &   $11.280$            &                       &                    &                     &                    \\
\hline
 $\Omega_{bbc}^{*}$             &$\{bb\}c$               &  $\frac{3}{2}^{+}$     &     1        &      0       &        $1^{+}$            &   $10.54\pm0.11$     &   $11.287$            &    $11.20$            &    $11.480$        &                     &   $11.19$          \\
 \hline
 $\Omega_{ccb}^{'}$             &$[cc]b$                 &  $\frac{1}{2}^{+}$     &     0        &      0       &        $0^{+}$            &   $7.49\pm0.10$      &                       &                       &                    &                     &                    \\
\hline
 $\Omega_{bbc}^{'}$             &$[bb]c$                 &  $\frac{1}{2}^{+}$     &     0        &      0       &        $0^{+}$            &   $10.35\pm0.07$     &                       &                       &                    &                     &                    \\
\hline\hline
\end{tabular}}
\label{table3}
\end{table}

\section{summary and outlook}\label{sec3}
In summary, we have studied the mass spectra of singly, doubly, and
triply heavy baryons in the framework of full QCD sum rules and
arrived at three conclusions in chief. First, our results for singly
heavy baryons are well compatible with the existing experimental
data. Second, the mass values for doubly heavy baryons are in
reasonable accord with other predictions. Third, the numerical
results for triply heavy baryons are lower than the predictions from
potential models, nevertheless, the one for $\Omega_{ccc}$ is in
good agreement with the lattice study.

Though enormous progress have been achieved in experimental and
theoretical aspects for the heavy flavor baryons, there are still
many problems desiderated to resolve. In experiment, it is worthy to
point out that most of the $J^{P}$ quantum numbers for the observed
heavy baryons have not been determined, but are assigned by PDG on
the basis of quark model predictions, which are looking forward to
further experimental identification, particularly for some higher
excited states. More data on singly bottom baryons and doubly heavy
baryons, along with the evidence on triply heavy baryons are
earnestly expected after the Large Hadron Collider startup, which
may supply a gap of experimental data in the future. Theoretically,
in order to improve on the accuracy of the QCD sum rule analysis for
the heavy baryons, especially for triply heavy baryons, one needs to
take into account the QCD $O(\alpha_s)$ corrections to the sum rules
in the further work. Additionally, it is interesting to carry out a
comprehensive study on triply heavy baryon spectra from lattice QCD
for the future.
\begin{acknowledgments}
J.~R.~Z. is very indebted to B.~L.~Ioffe and Yuan-Ben Dai for the
current communications and helpful discussions, and J.~R.~Z. is also
very grateful to M.~Nielsen for helpful communications on doubly
heavy baryons. This work was supported in part by the National
Natural Science Foundation of China under Contract No.10675167.
\end{acknowledgments}

\end{document}